\documentclass[final]{aipproc}

\usepackage{amsmath}
\usepackage{amssymb}

\layoutstyle{8x11double}

\begin{document}

\title{Acoustic Neutrino Detection in Ice: Past, Present, and Future}

\author{Timo Karg}{
  address={DESY, Platanenallee 6, 15738 Zeuthen, Germany},
  email={timo.karg@desy.de}
}

\begin{abstract}
  Acoustic neutrino detection is a promising technique to instrument
  the large volumes required to measure the small expected flux of
  ultra-high energy cosmogenic neutrinos. Using ice as detection
  medium allows for coincident detection of neutrino interactions with
  acoustic sensors, radio antennas and optical light sensors with the
  benefit of cross calibration possibilities or independent
  measurements of the the same event. We review the past development
  of the field and discuss its current status and challenges. Results
  from site exploration studies, mainly by the South Pole Acoustic
  Test Setup (SPATS) which has been codeployed with the IceCube
  neutrino telescope at South Pole, and current physics results are
  presented. Current ideas for the design, calibration, and deployment
  of acoustic sensors for new projects are shown. The possible role of
  the acoustic technique in future in-ice neutrino detectors is
  discussed.
\end{abstract}



\maketitle

\section{Introduction}

In the year 2012 we celebrate the $100^\mathrm{th}$ anniversary of the
discovery of cosmic rays. But even after a hundred years of research
many questions about the origin, acceleration, and composition of
ultra-high energy cosmic rays remain unanswered. The multi-messenger
approach, combining the information gained from electromagnetic
radiation from radio to TeV photons, charged cosmic rays, and
neutrinos promises to resolve these problems. Neutrinos are ideal
messengers in the sense that they are undeflected by magnetic fields
during their propagation and that they rarely interact, preserving
their initial direction and energy until detected at Earth.

Ultra-high energy (UHE; $E_\nu \gtrsim 100$~PeV) neutrinos, offer a
very rich physics program, including astrophysics, cosmology, particle
physics, and physics beyond the Standard Model:

\begin{itemize}
\item Cosmogenic neutrinos are produced in the interactions of charged
  cosmic rays at ultra-high energies with the cosmic microwave
  background \cite{Berezinsky:1969}, typically within a distance of a
  few ten Mpc of the source \cite{Greisen:1966}. Thus, for very far
  sources, they allow for a good pointing towards the source. The flux
  of UHE cosmogenic neutrinos is very sensitive to the chemical
  composition of the charged cosmic rays (e.g.~\cite{Kotera:2010,
    Ahlers:2012}).
\item Since UHE neutrinos reach us from very high redshifts, their
  flux is also sensitive to the evolution of cosmic ray sources in the
  earlier universe (e.g.~\cite{Kotera:2010}). Resonant $Z$ boson
  production could reveal the cosmic neutrino background and allow us
  to determine the neutrino masses \cite{Ringwald:2008}.
\item Measuring the neutrino flux with different mass overburden,
  e.g. at different zenith angles with an underground detector, will
  allow us to determine the neutrino absorption in the Earth and thus
  probe the neutrino nucleon cross section at high center-of-mass
  energies \cite{Connoly:2006}.
\item There are many theoretical models of physics beyond the Standard
  Model which predict large deviations of the neutrino nucleon cross
  section from the Standard Model at high energies
  (e.g.~\cite{Anchordoqui:2003, Mattingly:2010}). UHE neutrinos will
  allow us to test many of these models.
\end{itemize}

To achieve all these goals we need to measure UHE neutrinos with
reasonable statistics and good energy and direction resolution. This
requires a detector with a volume $\ge 100$~km$^3$. There are
different experimental techniques to build such large scale detectors
which are currently pursued either in running experiments or as
feasibility studies.

Radio detection experiments are looking for short radio pulses in the
hundreds of MHz to GHz frequency range emitted by the electromagnetic
cascade generated in a neutrino interaction. These experiments can be
embedded in radio-transparent, homogeneous media like ice
\cite{Besson:2012} or salt, or use balloons or satellites to observe
large natural ice volumes. Also, observations of the Moon with radio
telescopes are employed to look for neutrino interactions in the lunar
regolith \cite{Seckel:2012}.

Extensive air shower experiments can detect UHE neutrinos either as
highly inclined, ``young'' air showers, where the primary neutrino has
penetrated deep into the atmosphere before interacting, or as up-going
air showers from Earth-skimming neutrinos. The HiRes detector
\cite{Abbasi:2008} and the Pierre Auger Observatory \cite{Abreu:2012}
have used these methods to set upper limits on the flux of UHE
neutrinos.

Finally, acoustic neutrino detectors are searching for ultrasonic
pressure pulses generated in the instantaneous heating and expansion
of the medium induced by electromagnetic and hadronic cascades. Water
\cite{Graf:2012}, ice, salt, and permafrost soil \cite{Nahnhauer:2008}
have been discussed as detection media. In this work we review the
development, status and perspectives of acoustic neutrino detection in
ice.

\section{A brief history of acoustic neutrino detection in ice}

Acoustic neutrino detection in liquids is based on the thermo-acoustic
model \cite{Askaryan:1977, Bowen:1977}: When a neutrino of any flavor
interacts via a charged- or neutral current interaction, a hadronic
and/or electromagnetic cascade develops at the interaction vertex,
which carries a significant amount of the neutrino energy. In a dense
medium this energy is dissipated in a volume of typically $10$~m in
length and a few centimeters in diameter. This leads to an
instantaneous heating of the cascade volume. The corresponding rapid
expansion of the volume propagates as an ultrasonic shock wave
perpendicular to the cascade axis and can be measured as a short
(i.e.~with a broad frequency spectrum), bipolar pressure pulse with a
duration of several ten microseconds. The details of the pulse depend
on the material properties of the medium and on the modeling of the
cascade energy deposition density.

The first ideas about acoustic detection of particles in liquids date
back to the 1950s \cite{Askaryan:1957}. It was then revived in the
1970s and studied in great detail in the context of the DUMAND
project, leading to detailed calculations of the expected acoustic
signals from the thermo-acoustic model \cite{Learned:1979,
  Askaryan:1979} and first measurements with a proton beam from an
accelerator dumped in water \cite{Sulak:1979}.

With the design and construction of the AMANDA optical {Cherenkov}
neutrino telescope at South Pole interest in acoustic neutrino
detection in ice began. Ice, in contrast to water, allows for the
propagation of longitudinal (pressure, p) sound waves and transverse
(shear, s) waves. The formalism of the thermo-acoustic model can be
expanded to the case of solid media and predicts the excitation of
mainly pressure waves by neutrino interactions
(cf.~e.g.~\cite{Salomon:2007}); shear waves can be generated at
impurities in the crystal structure of the medium.

\begin{figure}
  \includegraphics[width=0.44\textwidth]{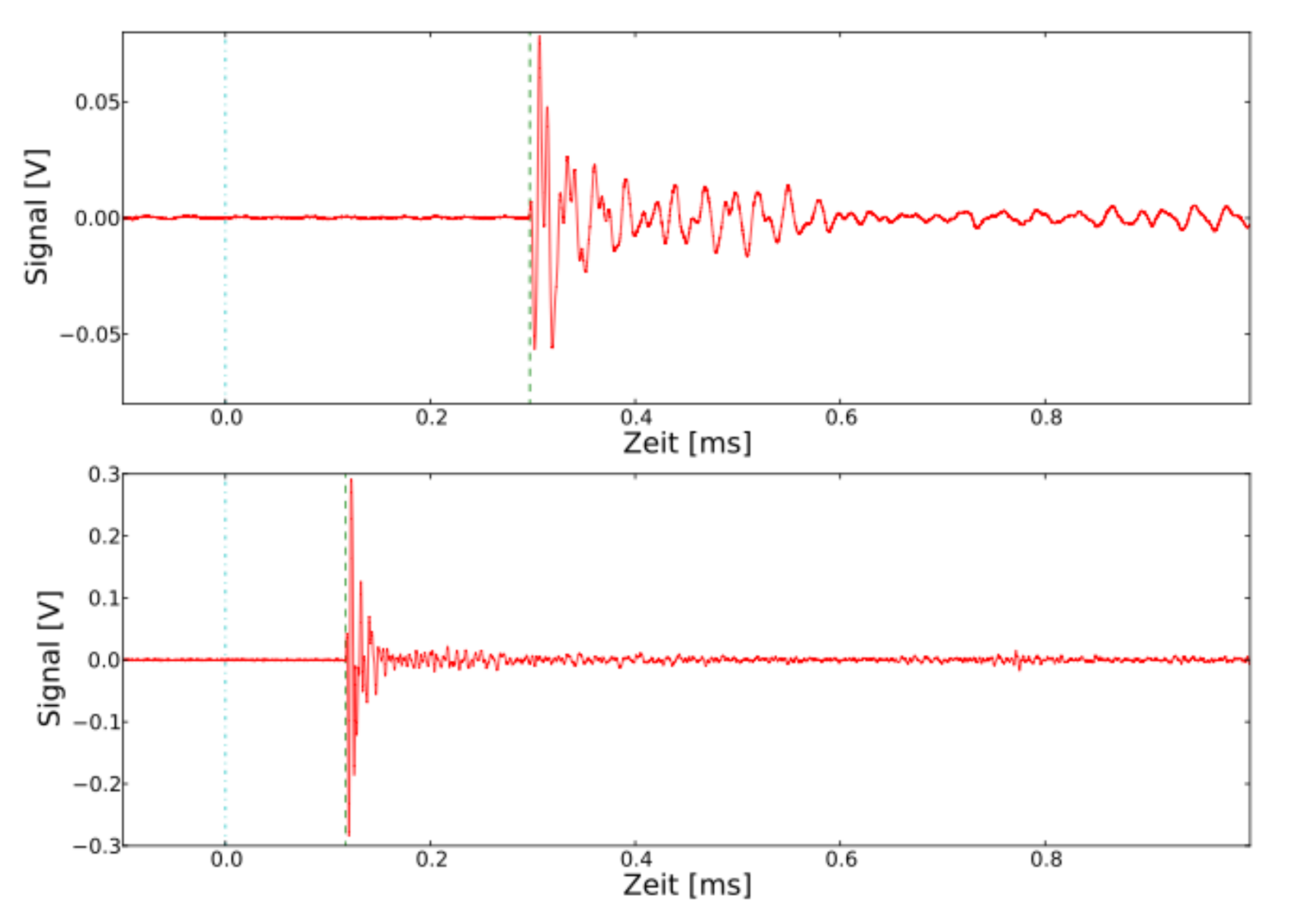}  
  \caption{Laser-induced acoustic pulses in water (top) and ice
    (bottom) generated with identical laser pulses (from
    \cite{Heinen:2011}).\label{fig:laser}}
\end{figure}

Acoustic signals predicted by the thermo-acoustic model scale, for
equal energy deposition densities, with the thermo-elastic properties
of the detection medium. Due to the nearly equal matter densities of
water and ice the energy deposition density from a neutrino
interaction is very similar. From detailed calculations, taking into
account the elastic properties of ice, the amplitudes of
neutrino-induced thermo-acoustic signals in ice are expected to have
amplitudes which are larger by a factor of about four compared to
water \cite{Salomon:2007}. This is supported by measurements in the
Aachen Acoustic Laboratory \cite{Heinen:2012} where laser-induced
acoustic pulses in water and ice have been studied
\cite{Heinen:2011}. Figure~\ref{fig:laser} shows that the scaling of the
signal amplitude from water to ice is compatible with expectations. It
can also be seen that in ice, due to the larger speed of sound, higher
frequency signals are generated.

Phenomenological studies of the ice acoustic properties predicted an
acoustic attenuation length of several kilometers \cite{Price:1993,
  Price:2006} and low background noise \cite{Price:2006}, which would
allow for very large, sparsely instrumented detection
volumes. Subsequent studies favored the radio technique as being more
sensitive than acoustics \cite{Price:1996}. In the same article
(Ref.~\cite{Price:1996}) the possibility of hybrid detection, using
several complementary techniques (radio and optical) is discussed. In
the following years experimental limits on the flux of high energy
neutrinos became more stringent and theoretical flux predictions
decreased accordingly. It became clear that detector volumes $\ge
100$~km$^3$ are required which are difficult to achieve with optical
Cherenkov detectors. Since radio and acoustic signals were expected to
have similar attenuation lengths, hybrid radio-acoustic detectors were
discussed and simulation studies showed very promising results
\cite{Vandenbroucke:2006}.

However, it was clear that the predicted acoustic properties of the
ice need to be tested by in-situ measurements. Different piezoelectric
sensors for use in ice were developed and characterized
\cite{Boser:2006} and sound generation in ice by an accelerator proton
beam was studied \cite{Nahnhauer:2003}. These efforts led to the
construction and deployment of the South Pole Acoustic Test Setup
(SPATS) that will be discussed later in this work.

\section{Measuring acoustic waves in ice}

Building a large acoustic detector with reasonable energy and
direction resolution requires to fully understand the sensor response
in-situ. Sensor sensitivity is not just a single number to convert
incident pressure to output voltage measured at the sensor, but is a
function of incident wave direction, wave mode, temperature, and
possibly other environmental parameters. Since no pre-calibrated
sensors for ice are commercially available that can be used for
relative calibration, extensive studies have been performed to use the
reciprocity calibration method, that does not require a reference
receiver, in ice \cite{Semburg:2011}. The in-situ calibration of
sensors deep in natural glacial ice is even more challenging due to
the limited possibilities of access to the detectors.

In the SPATS project (cf.~next section) it has been tried to factorize
the problem in the laboratory: SPATS sensors have been absolutely
calibrated in water at $0^\circ$C before deployment \cite{Abdou:2011}
and the angular response of the sensor has been determined at
different frequencies \cite{Fischer:2006}. It is not obvious whether the
calibration results can be transferred to operation conditions, where
the sensors are frozen in the deep ice at
South Pole. There, they are subject to low temperatures of
approx.~$-50^\circ$C, increased static pressure, and a different
sensor-medium interface (ice to steel). The different effects have
been studied separately in the laboratory:

\begin{figure}
  \includegraphics[width=0.44\textwidth]{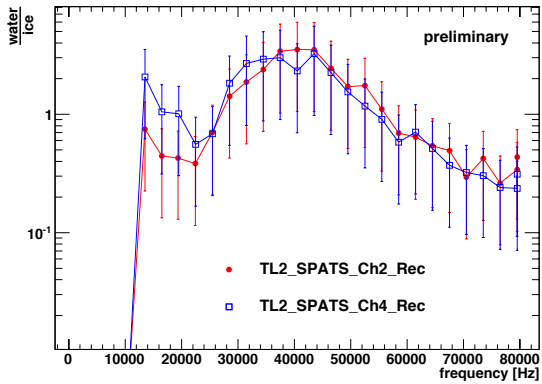}
  \caption{Ratio of the sensitivities measured in water and ice of two
    SPATS sensor channels (from
    \cite{Paul:2011}).\label{fig:sensitivity}}
\end{figure}

\begin{itemize}
\item In air at constant pressure, a sensor has been cooled down from
  $0^\circ$C to $-50^\circ$C and its response to the signal from an
  external transmitter, kept at constant temperature, has been used as
  an estimator for the receiver sensitivity. It has been found that
  the sensor sensitivity increases by a factor of $1.5 \pm 0.2$ when
  the temperature was lowered from $0^\circ$C to $-50^\circ$C
  \cite{Abdou:2011}.
\item In a pressure vessel, filled with an emulsion of water and oil,
  a sensor has been exposed to static pressure up to $100$~bar. A
  transmitter placed outside the pressure vessel and transmitting
  through the steel vessel has been used to determine possible changes
  in sensitivity. The results indicate that the variation of the
  receiver sensitivity is less than $30$\% between $1$ and $100$~bar
  static pressure \cite{Abdou:2011}.
\item The Aachen Acoustic Laboratory \cite{Heinen:2012} allows for the
  production of volumes up to $3$~m$^3$ of clear ice with temperatures
  down to $-25^\circ$C which are ideal for the study of changes in
  sensor sensitivity when going from the water to the ice
  phase. Figure~\ref{fig:sensitivity} shows the ratio of the
  sensitivities of a SPATS sensor measured in water and ice using the
  reciprocity calibration method. It can be seen that the ratio is
  compatible with unity within its errors, indicating that the
  sensitivity does not change in the frequency range relevant for
  acoustic neutrino detection (approx.~from $10$ to $50$~kHz) when the
  sensor is frozen into bulk ice.
\end{itemize}

Assuming that the influences of the environmental effects are
independent, it has been concluded that for the SPATS sensors the
sensitivity in ice will be increased by a factor of $1.5 \pm 0.4$
compared to the pre-deployment calibration in water
\cite{Abbasi:2012}. This factor takes into account the uncertainties
from the temperature and pressure measurements.

It is under investigation how naturally occurring transient noise
events and artificial calibration transmitters can be utilized for
sensor relative calibration and angular response measurements in-situ
\cite{Berdermann:2012}.

\section{Site exploration}

\begin{figure}
  \includegraphics[width=0.65\textwidth]{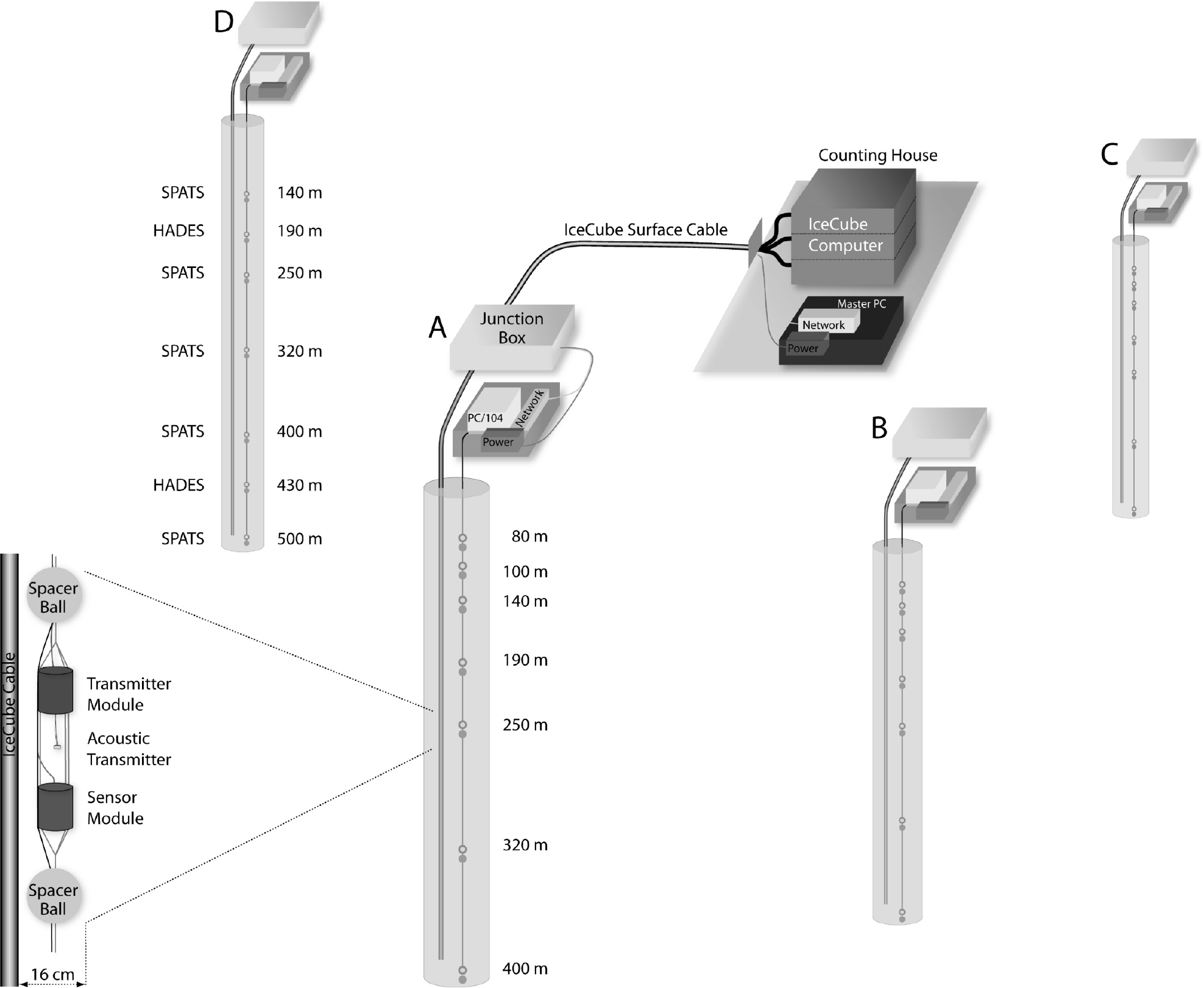}
  \caption{Overview of the South Pole Acoustic Test Setup (SPATS)
    frozen into the upper 500~m of holes drilled for the IceCube
    neutrino telescope (from \cite{Abdou:2011}).\label{fig:spats_overview}}
\end{figure}

Another important step towards a large scale detector embedded in a
natural detection medium is the full understanding of the signal
propagation properties and backgrounds therein. For an acoustic
experiment this means the determination of the sound speed depth
profile, the attenuation length, the noise level, and possible
transient backgrounds. The sound speed profile determines possible
refraction during the signal propagation that impedes accurate vertex
reconstruction up the existence of multiple solutions. The attenuation
length and noise level will determine the detector geometry required
to achieve a given neutrino energy threshold. Transient noise sources
need to be identified and characterized to separate them from neutrino
induced events. The measurement of the sound speed profile and
transient backgrounds are easier to accomplish in the sense that they
only rely on time information. The attenuation length and noise level
measurement depend on amplitude information and are thus subject to
the calibration challenge discussed in the previous section.

\subsection{SPATS -- Hardware}

To carry out these measurements in the Antarctic ice at the South
Pole, the site of the IceCube neutrino observatory, the South Pole
Acoustic Test Setup (SPATS) has been designed and is successfully
operated since January 2007 \cite{Abdou:2011}. SPATS consists of four
vertical strings that are deployed in the upper $500$~m part of
IceCube bore holes after the installation of the optical IceCube
string. Horizontal baselines between $125$~m and $543$~m are
covered. Each SPATS string is instrumented with seven stages, each
containing an acoustic receiver and a transmitter. The SPATS sensor is
made from a steel housing with three piezoceramic disks pressed to the
inner wall at $120^\circ$ separation for full azimuthal coverage. The
signals are amplified in the sensor module and the differential
analogue signal is transmitted via twisted pair cable to the surface
where it is digitized and time stamped in a String-PC. The data from
all four strings are collected by a Master-PC housed in the IceCube
counting house and are prepared for satellite transmission to the
IceCube central data storage. At two positions an alternative sensor
type, HADES, is installed, where the piezoceramic element is cast in
resin and mounted below the steel housing. HADES is used for
systematic studies of the sensor medium coupling. The SPATS
transmitter consists of a piezoceramic ring cast in resin and frozen
directly into the ice. It is connected to a high voltage pulser which
is protected in a steel housing and steered by the String-PC. A
schematic overview of the SPATS hardware is shown in
Fig.~\ref{fig:spats_overview}.

SPATS is complemented by a mobile acoustic transmitter, called
``pinger'', which can be lowered into freshly drilled water filled
IceCube holes while continuously emitting acoustic pulses with high
stability. The pinger is retrieved from the hole after operation.

\subsection{SPATS -- Results}

\begin{figure}
  \includegraphics[width=0.5\textwidth]{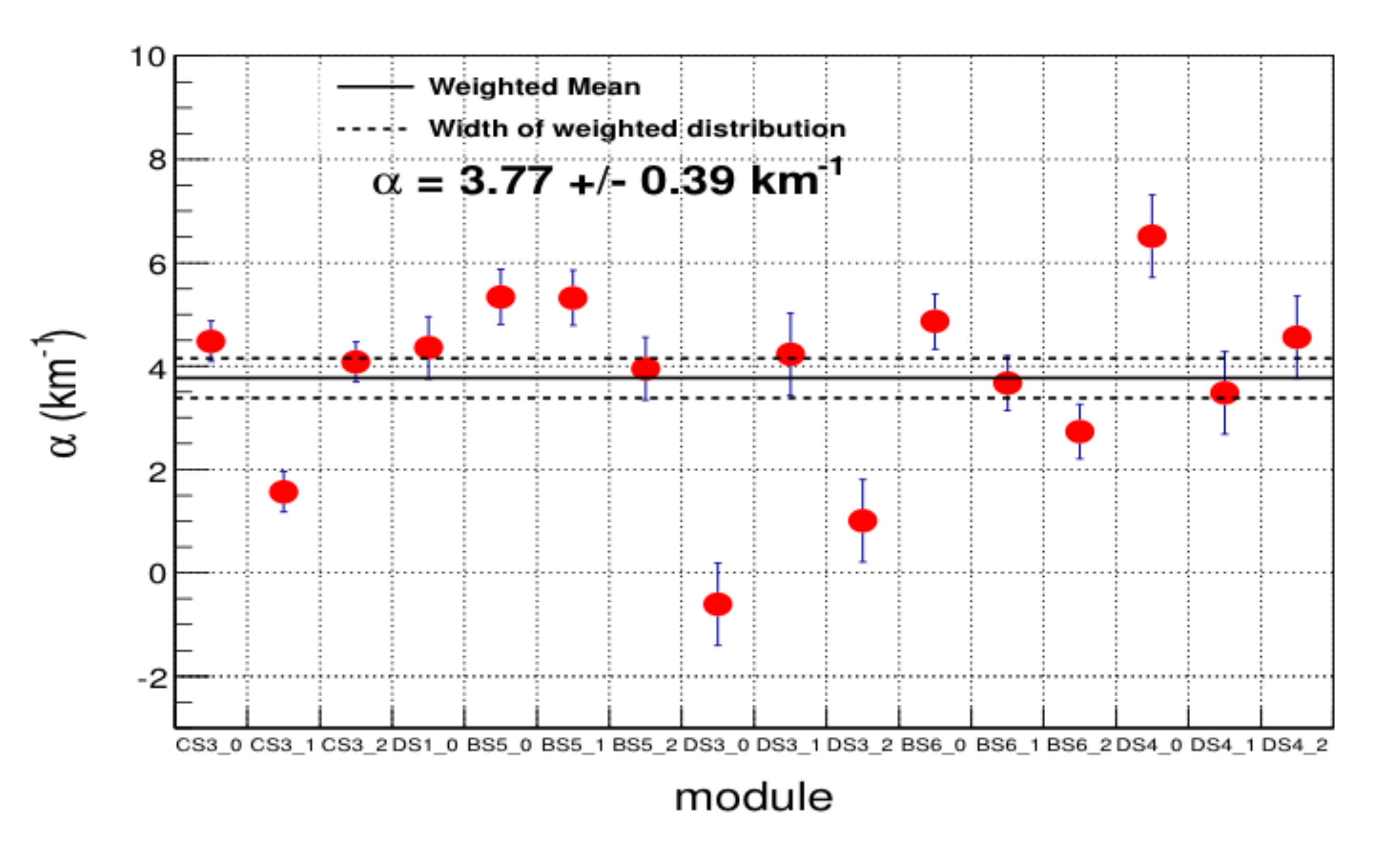}
  \hspace{0.5cm} \includegraphics[width=0.5\textwidth]{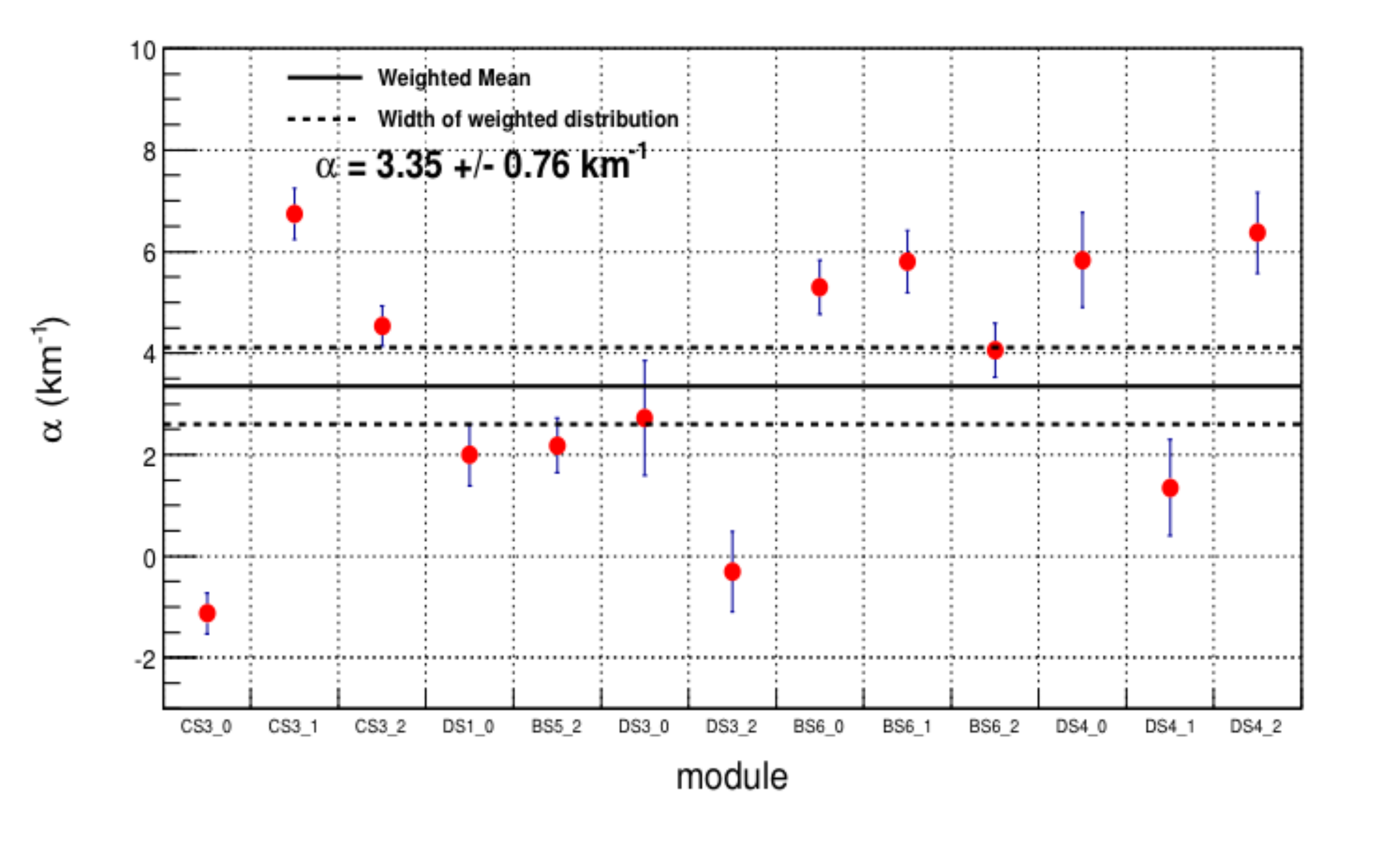}
  \caption{Acoustic attenuation coefficient for 30~kHz (left) and
    45~kHz (right) measured with the SPATS retrievable pinger in
    different sensor channels (from
    \cite{Abdou:2012}).\label{fig:attenuation}}
\end{figure}

SPATS has measured the speed of sound depth profile in the Antarctic
ice at the South Pole in the depth range from $80$~m to $500$~m using
the retrievable pinger over horizontal baselines of $125$~m
\cite{Abbasi:2010}. The pinger emits longitudinal waves that propagate
through the water column in the drill hole and are then transmitted
into the ice. When the incidence on the water-ice interface is
non-normal, part of the waves energy is transferred into a shear
wave. Thus, the sound speed profile for pressure and shear waves could
be determined. The speed of sound is found to be increasing in the top
$200$~m of the ice where a gradual transition from a snow/air mixture
occurs (firn layer) and is found to be constant below that depth. The best
fit values \cite{Abbasi:2010} for the sound speed $v$ at $375$~m depth
and its gradient $g$ at this depth are for pressure waves

\begin{eqnarray*}
  v_p & = & (3878 \pm 12) \, \mathrm{m} \, \mathrm{s}^{-1} \\
  g_p & = & (0.087 \pm 0.13) \, \frac{\mathrm{m} \,
    \mathrm{s}^{-1}}{\mathrm{m}}
\end{eqnarray*}

\noindent and for shear waves 

\begin{eqnarray*}
  v_s & = & (1975.8 \pm 8.0) \, \mathrm{m} \, \mathrm{s}^{-1} \\
  g_s & = & (0.067 \pm 0.086) \, \frac{\mathrm{m} \,
    \mathrm{s}^{-1}}{\mathrm{m}}
\end{eqnarray*}

\noindent Since the gradient is compatible with zero only very little
refraction is expected below the firn layer.

Pinger data have also proven very valuable to determine the signal
amplitude attenuation length \cite{Abbasi:2011} in the frequency range
from $10$ to $30$~kHz. This analysis requires the comparison of the
signal observed at different distances. To reduce systematic
uncertainties from the sensor absolute calibration and angular
response, attenuation lengths are derived for each sensor channel. For
this the pinger, which produces highly reproducible pulses, is
deployed in different IceCube drill holes at increasing distances, but
aligned in direction, so that only a small range of the angular
response of the sensor is probed. Averaging over all available sensor
channels leads to an signal amplitude attenuation length
\cite{Abbasi:2011} of

\begin{equation*}
\langle \lambda \rangle = 312^{+68}_{-47} \, \mathrm{m}
\end{equation*}

\noindent This value is much smaller than the several kilometers
initially expected from theoretical calculations \cite{Price:2006}. To
study this discrepancy a modified pinger has been constructed that
emits gated sine bursts at different frequencies ($30$, $45$, and
$60$~kHz). It has been used to study the frequency dependence of the
attenuation length. Absorption, if it would be the main attenuation
mechanism, is expected to be frequency independent, whereas the
scattering coefficient would increase $\propto f^4$, where $f$ is the
frequency of the signal \cite{Price:2006}. The data from the modified
pinger are analyzed with the same methods as described above
\cite{Abdou:2012}. Figure~\ref{fig:attenuation} shows the attenuation
coefficient measured in the different sensor channels for $30$~kHz and
$45$~kHz as well as the mean value and spread of the data points. The
signal-to-noise ratio for $60$~kHz has been too poor to extract an
attenuation length since the transmit response of the pinger
piezoelectric element is small at this frequency. It can be seen that
the two values are compatible with a frequency independent attenuation
length and that an $f^4$ frequency dependence is hard to reconcile
with the data.

SPATS unbiased noise data, that are recorded every hour for $0.1$~s
have been used to study the background noise level. It has been shown
that the noise is Gaussian and that the RMS is very stable over time
\cite{Abbasi:2012}. Using the corrections on the sensor calibration
discussed in the previous section, the absolute noise level below the
firn layer has been estimated to be $14$~mPa, corresponding to the
signal expected from a $10^{11}$~GeV neutrino interacting at a
distance of $1000$~m to the sensor \cite{Abbasi:2012}. However, the
systematic uncertainty on the noise level is still large, making
additional measurements with sensors pre-calibrated in ice desirable.

\begin{figure}
  \includegraphics[width=0.475\textwidth]{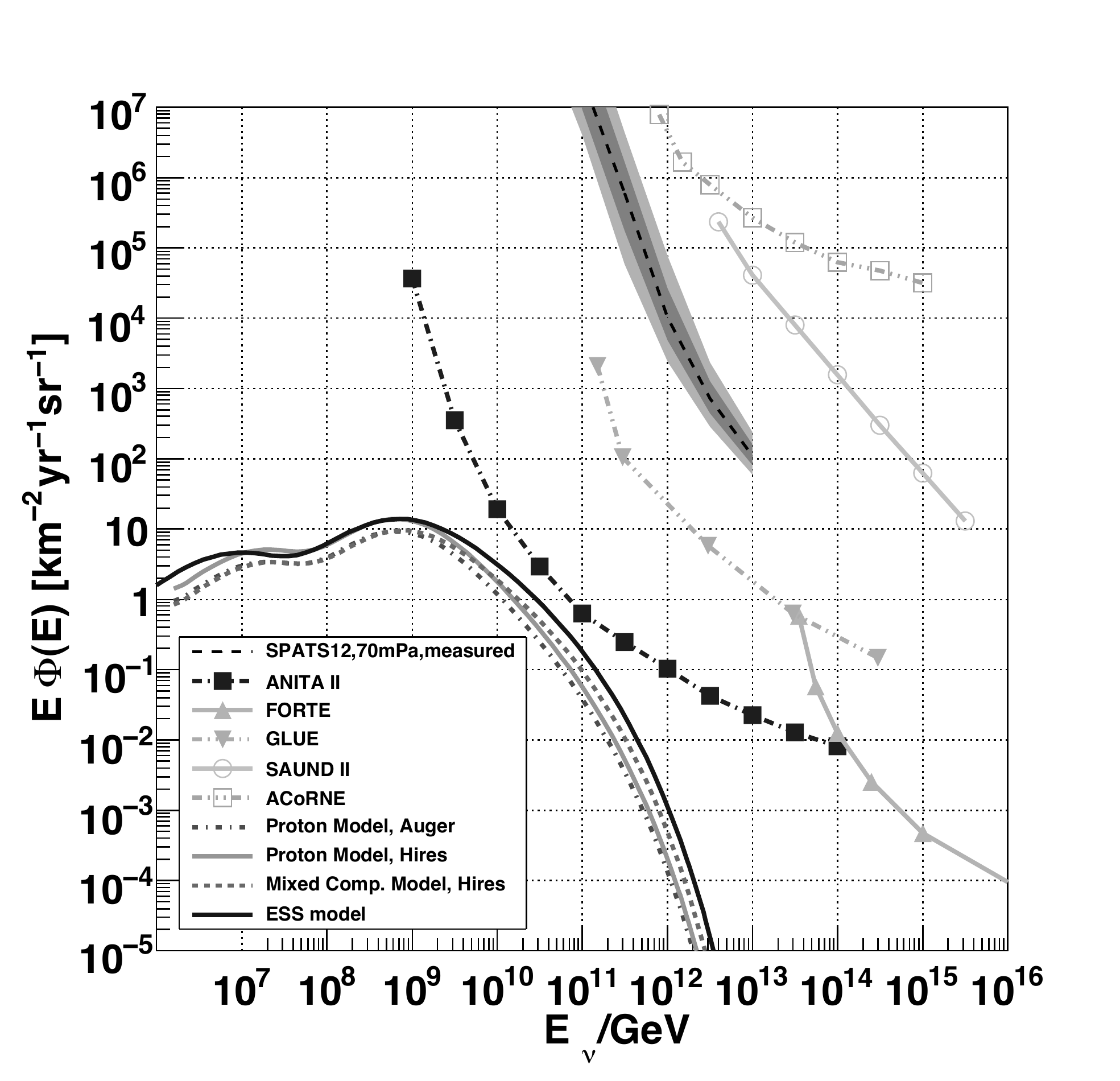}  
  \caption{Upper limits (90\% C.L.) on the all-flavor flux of
    ultra-high energy neutrinos set by acoustic neutrino detection
    experiments in water (SAUND II, ACoRNE) and ice (SPATS12). Several
    models on for the flux of cosmogenic neutrinos and the most
    stringent limits set by radio experiments are shown for comparison
    (adapted from \cite{Abbasi:2012}).\label{fig:limits}}
\end{figure}

An analysis of transient events triggering all four strings of the
SPATS detector revealed only man-made sources: all events were
reconstructed to either re-freezing IceCube drill holes or Rodriguez
Wells, large caverns in the ice used to circulate the water for the
IceCube hot water drill system \cite{Abbasi:2012}. The absence of
unidentified transient noise sources in the deep ice allowed setting
an upper limit on the flux of ultra-high energy
neutrinos. Figure~\ref{fig:limits} compares the limit set by SPATS to
other experiments and the expected cosmogenic neutrino flux. It has to
be kept in mind that none of the acoustic experiments shown there were
designed as neutrino detectors, but were built for site-exploration
(SPATS) or parasitically use existing military hydrophone arrays to
search for neutrino induced signals (SAUND II, ACoRNE).

\section{A path forwards}

The SPATS results show that acoustic neutrino detection in the glacial
ice at the South Pole is feasible. The acoustic attenuation length is
shorter than for radio signals
but of comparable magnitude which opens the possibility to design a
hybrid radio/acoustic detector. Due to the intrinsically higher energy
threshold of the acoustic technique a possible scenario is to use the
radio sub-array to trigger the acoustic sub-array up to energies where
the acoustic detector becomes fully efficient by itself. In this case
a single in-time hit in an acoustic sensor can already be highly
significant since all known backgrounds produce either only radio
emission or only sound emission.

The construction of such a detector at the South Pole will require the
deployment of a few hundred strings per $100$~km$^2$ instrumented area
(estimating a spacing of $500$~m between strings based on the measured
attenuation lengths for radio and acoustic) reaching below the firn
layer. A design which is scalable in size is desirable so that, once
the magnitude of the flux of ultra-high energy neutrinos is measured,
the detector can be expanded in size to accumulate sufficient event
statistics to answer the physics questions discussed in the
introductory section.

\begin{figure}
  \includegraphics[width=0.35\textwidth]{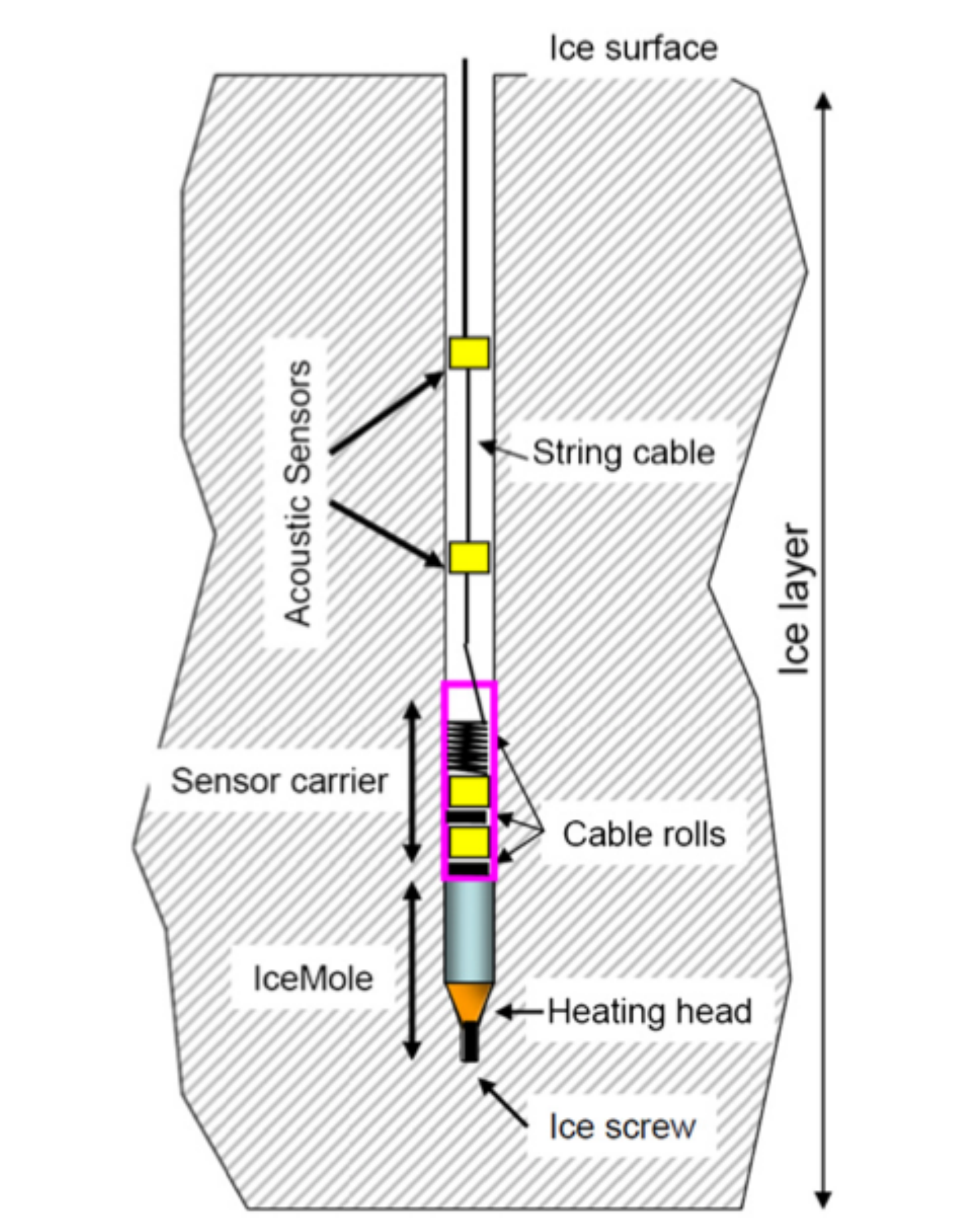}  
  \caption{Schematic of an autonomous drilling and deployment probe
    based on the IceMole concept (cf.~text). The cable and sensors are
    stored in and deployed from the probe, so that the hole is allowed
    to refreeze immediately after the passage of the probe (from
    \cite{Laihem:2012}).\label{fig:drilling}}
\end{figure}

A strong R\&D program will be required to realize a new large area
hybrid detector. Apart from South Pole there are several other sites
in Antarctica where scientific infrastructure exists and which are
worth evaluating:\\
- On the Ross Ice Shelf the ARIANNA radio neutrino
detector\footnote{http://arianna.ps.uci.edu/} is currently under
construction.\\
- Concordia Station at Dome C is located on top of more than $3$~km of
very cold ice which is favorable for acoustic and radio signal
propagation.

The installation of a new detector consisting of several hundred to a
few thousand strings will also require new techniques for drilling and
deployment. One possibility that is being discussed is the use of
autonomous drilling- and/or melting-probes similar to the IceMole
\cite{Dachwald:2011} prototype. To minimize human intervention in the
drilling and deployment procedure, the sensors and cable would be
stored in and deployed from the disposable icecraft which would remain
in the ice at the bottom of the string after deployment. This allows
for the hole to refreeze immediately after the passage of the probe. A
schematic of the procedure is shown in Fig.~\ref{fig:drilling}.

A large area detector at a remote site will also require new concepts
for calibration, communication and power supply. The large overall
extent and the large distances between the components prohibit a fully
cabled design. Wireless data transmission and power generation at the
site of the component by e.g.~wind or solar power are
required. Ideally one would have a combination of wind and solar power
since beyond the polar circles solar power is unavailable half of the
year. Valuable lessons can be learned from large area experiments
which are already in operation, like the Pierre Auger Observatory
\cite{Abraham:2004}.

\section{Conclusions}

Ultra-high energy neutrinos offer a vast physics program covering
astrophysics, cosmology, particle physics, and physics beyond the
Standard Model. The acoustic neutrino detection technique has made
large advances over the last few years: sensors have been designed and
their behavior in ice is largely understood. The acoustic properties
of the South Pole glacier have been measured and found to be suitable
for neutrino detection. It is expected that acoustic can play an
important part in a future hybrid neutrino telescope in ice. To
realize this a strong R\&D program has to be established and first
promising studies have already been presented.

\begin{theacknowledgments}
  T.K.~is supported by the ``Helmholtz Alliance for Astroparticle
  Physics HAP'' funded by the Initiative and Networking Fund of the
  Helmholtz Association.
\end{theacknowledgments}

\bibliographystyle{aipproc}
\bibliography{arena2012_karg}
\end{document}